\documentclass[11pt,a4paper]{article}
\pdfoutput=1
\usepackage{jheppub}
\usepackage{hyperref}
\usepackage[english]{babel}
\usepackage{graphicx}
\usepackage{bm}
\usepackage{slashed}
\usepackage{color}
\usepackage{multirow}
\usepackage{array}
\usepackage{hhline}

\DeclareFontFamily{T1}{pzc}{}
\DeclareFontShape{T1}{pzc}{m}{it}{<-> s * [1.10] pzcmi7t}{}
\DeclareMathAlphabet{\mathpzc}{T1}{pzc}{m}{it}

\title{CP Violation with Majorana neutrinos in $K$ Meson Decays.}

\author{ Claudio O. Dib$^{1,}$\footnote{claudio.dib@usm.cl},
Miguel Campos$^{1}$\footnote{miguel.campos@postgrado.usm.cl},
\ and  C. S. Kim$^{2,}$\footnote{cskim@yonsei.ac.kr} }
\affiliation{ $^{1}$ Centro Cient\'\i fico Tecnol\'ogico de Valpara\'\i so
and  Department of Physics, Universidad T\'ecnica Federico Santa
Mar\'\i a, Valpara\'\i so, Chile \\
$ ^{2}$ Department of Physics and IPAP, Yonsei University, Seoul 120-749, Korea }

\abstract{
\noindent We study the possibility of having CP asymmetries in the decay $K^\pm\to \pi^\mp\ell^\pm\ell^\pm$ $(\ell=e,\mu)$. This decay violates Lepton Number by two units and occurs only if there are Majorana particles that mediate the transition.
Even though the absolute rate is highly suppressed by current bounds, we search for Majorana neutrino scenarios where the CP asymmetry arising from the lepton sector could be sizeable. This is indeed the case if there are two or more Majorana neutrinos with similar masses in the range around $10^2$ MeV. In particular, the asymmetry is potentially near unity if two neutrinos are nearly degenerate, in the sense $\Delta m_N \sim \Gamma_N$. The full decay, however, may be difficult to detect not only because of the suppression caused by the heavy-to-light lepton mixing, but also because of the long lifetime of the heavy neutrino, which would induce large space separation between the two vertices where the charge leptons are produced. This particular problem should be less serious in heavier meson decays, as they involve heavier neutrinos with shorter lifetimes.
}
\date{\today }

\begin{document}

\maketitle


\section{Introduction}
\label{sec:Intro}

As it is now well established, neutrinos are not massless particles, opening the possibility of having CP violation originated in the leptonic sector  \cite{Branco:2011zb}.  This phenomenon can arise in a similar way as in the quark sector, namely complex phases in the mixing matrix.
In the leptonic sector this mixing matrix is called the Pontecorvo-Maki-Nakagawa-Sakata matrix ($U_{\rm PMNS}$) \cite{pmns}. From solar and atmospheric  neutrino experiments, 
we  have known for about a decade that the mixing angles $\theta_{12}$ and $\theta_{23}$ are rather large \cite{Angles}. Recently, the last of the mixing angles, $\theta_{13}$ was shown to be non zero as well \cite{theta-13}.
Consequently, the presence of at least one complex phase can not be ruled out.
Moreover, being neutrinos electrically neutral particles, its Lagrangian allows a Majorana mass term in addition to the Dirac mass terms that are allowed for all other fermions in the SM. In such case, two additional CP phases can arise and so CP violating processes appearing in the leptonic sector are not necessarily all inter-dependent. The Majorana character of neutrinos is also relevant in the explanation of the smallness of neutrino masses. At the theoretical level these masses are considered unnaturally small, unless they are explained by seesaw mechanisms \cite{see-saw}, where some of the neutrinos become light just as others become heavy. The latter are usually sterile under the SM interactions, interacting with SM particles only through lepton mixing.  To date many  scenarios of seesaw models with extra neutrinos have been proposed, with great variety of masses, from a few keV up to GUT scale near $10^{15}$ GeV \cite{seesawmodels}.  A particularly interesting model that is potentially relevant for the existence of leptonic CP at energy scales accessible to current experiments considers in the spectrum a pair of nearly degenerate neutrinos \cite{nuMSM}. 
Please note that the model, $\nu$MSM~\cite{nuMSM},
proposes two almost degenerate Majorana neutrinos with mass
between $100$ MeV and a few GeV, in addition to a light Majorana
neutrino of mass $\sim 10$ keV.
This almost degeneracy opens the possibility of having interfering amplitudes in meson decays where neutrinos are in s-channel intermediate states. Due to the near degeneracy, the interference provides an absorptive phase that is necessary for a {CP} asymmetry to appear. The experimental observation of the processes are clearly challenging, but the mechanism is worth studying.

Here we focus on the decay $K^\pm\to\pi^\mp\ell^\pm\ell^\pm$ ($\ell=e,\mu$), but the general procedure can be applied to decays of heavier mesons as well \cite{kim-cvetic}. This decay can only be mediated by Majorana particles, so its sole observation would be evidence of  their Majorana character.  In addition, an asymmetry in the $K^+$ and $K^-$ decay rates would be evidence of leptonic {CP} violation. This asymmetry could be sizeable if the process includes interfering amplitudes mediated by different neutrinos that are almost degenerate in mass, in the range between $m_\pi$ and $m_K$.
For decays of heavier mesons, the required neutrino masses would lie on a correspondingly higher mass range.


In Section II we describe the theoretical scenario for leptonic CP violation we are considering, specifying the formulation of the processes and studying the conditions for the leptonic CP violation to arise.
In Section III we present our results on the decay rates and the issues of observability, especially considering the long lifetimes of the intermediate neutrinos. Conclusions with a brief summary of the results are given in Section IV. Some details of the decay rate calculations related to the CP asymmetry are collected in an Appendix.


\section{The Scenario for Leptonic CP Violation.}
\label{sec:Scenario}

As stated above, we want to study $\Delta L=2$ charged meson decays mediated by heavy Majorana neutrinos with the prospect of obtaining a signal of leptonic CP violation. Here we focus on the charged kaon decay $K^{\pm} \to \ell^{\pm}\ell^{\pm}\pi^{\mp}$, but our conclusions can be extended to the decay of heavier mesons. The CP violation signal we seek in this kind of processes is simply the asymmetry in the decays of $K^+$ vs. $K^-$:
\begin{equation}
A_{CP} = \frac{\Gamma(K^+\to \pi^-\ell^+\ell^+)-\Gamma(K^-\to\pi^+ \ell^-\ell^-)}
{\Gamma(K^+\to \pi^-\ell^+\ell^+)+\Gamma(K^-\to\pi^+ \ell^-\ell^-)}.
\end{equation}

As CP violation arises from complex phases in the transition amplitudes, an observable effect only appears due to  interference of two or more amplitudes. In the case of charged particle decays, as it is well known \cite{deGouvea:2002gf}, the interfering amplitudes must have different  CP-odd phases $\phi_i$ ($\phi_i-\phi_j\neq 0$ mod $\pi$) and different CP-even phases $\eta_i$ ($\eta_i-\eta_j\neq 0$ mod
$\pi$). The CP-odd phases are those that come from the Lagrangian of the theory, in our case from the mixing in the lepton sector. These phases change sign between a process and its conjugate. The CP-even phases appear as absorptive parts in the transition amplitudes and do not change sign for the conjugate process. Our case of interest is when the amplitudes are mediated by two nearly degenerate neutrinos, with masses $m_{N_i}$ ($i=1,2$) in the range $m_\pi < m_{N_i}< m_K$.

\begin{figure}[t]
\begin{center}
\includegraphics[scale=0.70]{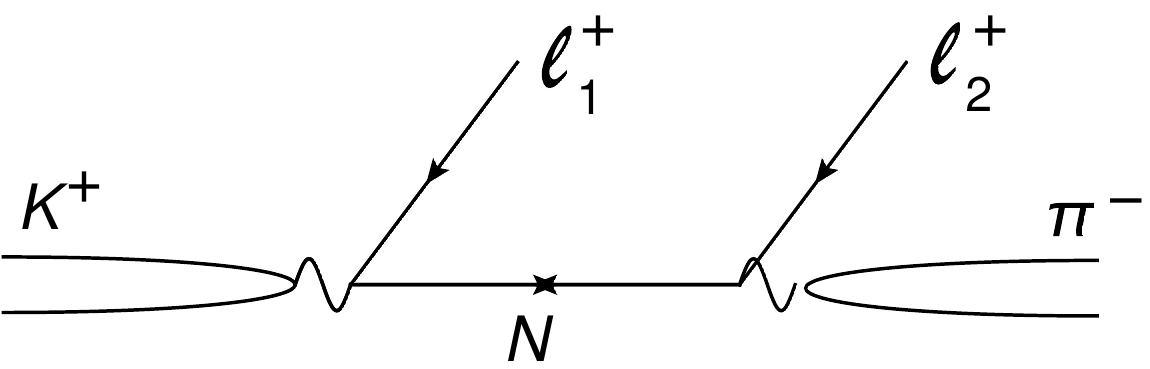}
\end{center}
	\caption{Diagram of the dominating amplitude for the lepton number violating decay $K^+\to \pi^- \ell^+\ell^+$ mediated by a Majorana neutrino $N$ with mass in the range between $m_\pi$ and $m_K$.}
	\label{fig:Kdecay-a}
\end{figure}

The dominant transition amplitudes mediated by these neutrinos \cite{Cvetic:2010rw} are diagrammatically shown in Fig.~\ref{fig:Kdecay-a}, and expressed as:
\begin{equation}
\mathcal{M}_{N_i}=- G_F^2  f_\pi f_K \   V^\ast _{ud} V^\ast _{us}
 \  \frac{\lambda_i \ U_{N_i\ell}^{2}\  m_{N_i}}{p_N^2-m_{N_i}^2+im_{N_i} \Gamma_{N_i}}    \bar{u}(l_2)\ \slashed{p} \  \slashed{k} \ (1-\gamma_5)v(l_1) ,
\label{amplitude1}
\end{equation}
where  $k$, $p$, $l_1$, $l_2$ and $p_N$ are the momenta of the kaon, pion, charged leptons, and intermediate neutrino, respectively, $U_{N_i\ell}$ is the lepton mixing matrix element and $\lambda_i$ the Majorana phase associated to $N_i$.
The squared of the total amplitude that includes both intermediate neutrinos is then:
\begin{equation}
 |\mathcal{M}_{N_1}+\mathcal{M}_{N_2}|^2 = |\mathcal{M}_{N_1}|^2 + |\mathcal{M}_{N_2}|^2 + 2 Re\left[ \mathcal{M}_{N_1} \mathcal{M}_{N_2}^\ast\right].\label{M_total}
\end{equation}
%
%

From here we can extract two common factors in this sum:  a constant factor
\[
 C_f  \equiv G_F^4  f_\pi ^2 f_K ^2 m_K^4\   |V_{ud}|^2 |V _{us}|^2 
 \]
and the lepton trace (summed over polarizations) $  |  \bar{u}(l_2)\ \slashed{p} \  \slashed{k} \ (1-\gamma_5)v(l_1)|^2 \equiv m_K^6  |L|^2$:
\begin{eqnarray}
|L|^2 &=& \frac{8}{m_K^6}\bigg\{
4(k\cdot p)(k\cdot\ell_1)( p\cdot\ell_2) + m_K^2 m_\pi^2 (\ell_1\cdot \ell_2) 
\nonumber\\
&&- 2 m_K^2 (p\cdot \ell_1)(p\cdot \ell_2) - 2 m_\pi^2 (k\cdot\ell_1)(k\cdot \ell_2)
 \bigg\}
 \label{trace}
\end{eqnarray}
Notice that we defined $C_f$ and $|L|^2$ to be dimensionless.
Accordingly, the first two terms in Eq.\ (\ref{M_total}) are similar:
\begin{equation}
|\mathcal{M}_{N_1}|^2+|\mathcal{M}_{N_2}|^2=  {C_f} |L|^2 m_K^2 \left\{ \frac{|U_{N_1 \ell}|^4 m_{N_1}^2}
{(p_N^2-m_{N_1}^2)^2+m_{N_1}^2\Gamma_{N_1}^2 }
+
 \frac{|U_{N_2 \ell}|^4 m_{N_2}^2}
{(p_N^2-m_{N_2}^2)^2+m_{N_2}^2\Gamma_{N_2}^2 } \right\} , \label{mag}
\end{equation}
while the interference term contains the phases that generate the CP asymmetry:
\begin{eqnarray}
2 Re\left[ \mathcal{M}_{N_1} \mathcal{M}_{N_2}^\ast\right]&=& 2 C_f |L|^2 m_K^2
\frac{|U_{N_1\ell}|^2 m_{N_1} }{\sqrt{(p_N^2-m_{N_1}^2)^2 +m_{N_1}^2\Gamma_{N_1}^2}}
\nonumber\\
&&
\quad \times \frac{|U_{N_2\ell}|^2 m_{N_2} }{\sqrt{(p_N^2-m_{N_2}^2)^2 +m_{N_2}^2\Gamma_{N_2}^2}}
\cos \left(  \Delta\eta -\Delta\phi   \right) .
\label{interference}
\end{eqnarray}
Here $\Delta\phi = \phi_2 -\phi_1$ are the CP-odd phases, namely $\phi_1$ is the phase of
$\lambda_1 U_{N_1\ell}^2$ and likewise for $\phi_2$, while $\Delta \eta = \eta_2 - \eta_1$
is the difference between the absorptive phases, defined as:
\begin{equation}
\tan \eta_1 = \frac{m_{N_1} \Gamma_{N_1}}{p_N^2-m_{N_1}^2} , \quad
\tan \eta_2 = \frac{m_{N_2} \Gamma_{N_2}}{p_N^2-m_{N_2}^2} .
\end{equation}

Since the neutrino resonances are very narrow, these absorptive phases are very small for all $p_N^2$ except near their poles at $m_{N_1}^2$ and  $m_{N_2}^2$, respectively. As we will show below, this feature causes the CP asymmetry to be sizeable if the two intermediate neutrinos
are nearly degenerate.

\begin{figure}[h!]
\begin{center}
\includegraphics[scale=0.70]{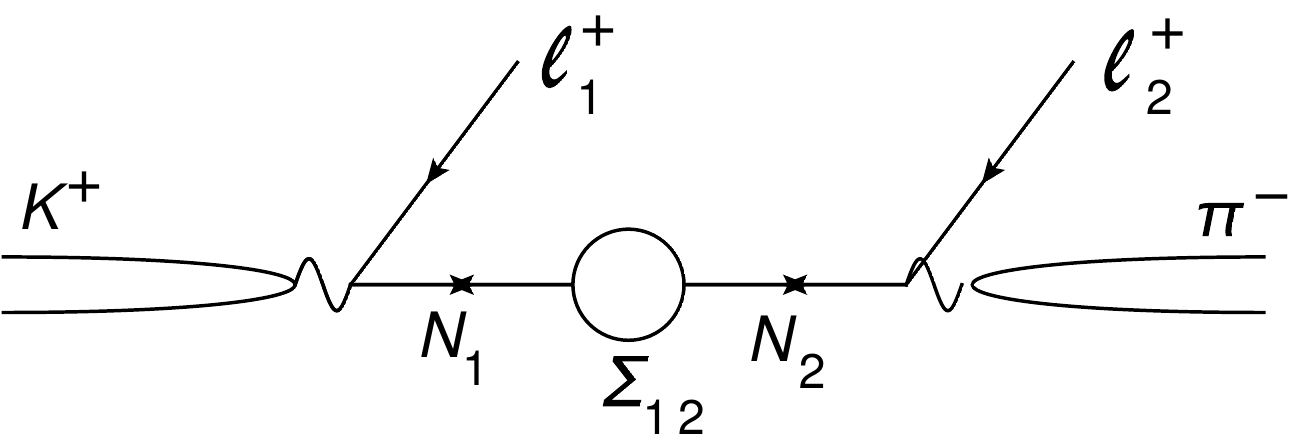}
\end{center}
	\caption{Non-diagonal contribution of the absorptive part of the 2-neutrino propagator to the amplitude for the lepton number violating decay $K^+\to \pi^- \ell^+\ell^+$.}
	\label{fig:Kdecay-2}
\end{figure}

There is just one caveat in the expressions above: we have disregarded the possibility that a diagonal $2\times 2$ neutrino mass matrix may not simultaneously diagonalize the absorptive part, i.e. it may remain non-diagonal elements of the 1-particle-irreducible (1PI) correction to the $2\times 2$ neutrino propagator, $\textrm{Im}\, \Sigma(p)_{12}$, as studied in Ref.\,\cite{Bray:2007ru}.  Such a contribution corresponds to additional diagrams of the form shown in Fig.\,\ref{fig:Kdecay-2}, where the 1PI correction $\Sigma_{12}$ stars at 1 loop and is assumed to be a purely absorptive part. In the 
notation of Ref.\,\cite{Bray:2007ru} the 1PI is parametrized as:
\[
\textrm{Im}\,\Sigma_{ij}(p) = A_{i j}(p^2)\not p P_L + A^\ast_{ij}(p^2) \not p P_R ,
\]
where $A^\ast_{ij} = A_{ji}$, for $i,j= N_1, N_2$. It is straightforward to show that the diagonal terms, $i=j$, correspond to the decay widths as 
$\Gamma_N = m_N (A_{NN} + A^\ast_{NN}) = 2 m_N A_{NN}$. In the non diagonal case, Fig.\,\ref{fig:Kdecay-2}, one can show that the relevant combination that enters is $(m_{N_1}  A_{12} + m_{N_2} A^\ast_{12} ) \equiv \Gamma_{12}$. Indeed, the amplitudes that follow from
Fig.\,\ref{fig:Kdecay-2}, analogous to Eq.~\eqref{amplitude1}, are:
\begin{eqnarray}
\mathcal{M}_{12} + \mathcal{M}_{21} &=& i G_F^2  f_\pi f_K \   V^\ast _{ud} V^\ast _{us}
 \ (\lambda_1 +\lambda_2)\ U_{N_1\ell} U_{N_2\ell}  \frac{\Gamma_{12}}{\Delta m^2} 
 \nonumber\\
&& \times  \left( \frac{m_{N_2}^2}{p_N^2-m_{N_2}^2 +i\epsilon} -
 \frac{m_{N_1}^2}{p_N^2-m_{N_1}^2+i\epsilon}
 \right)
    \bar{u}(l_2)\ \slashed{p} \  \slashed{k} \ (1-\gamma_5)v(l_1) ,
\label{amplitude11}
\end{eqnarray}
where $\Delta m^2 \equiv m_{N_2}^2 - m_{N_1}^2$.
Consequently, the inclusion of the non-diagonal amplitudes of Eq.~\eqref{amplitude11} in Eq.~\eqref{M_total} and the expressions thereafter is essentially the replacements:
\begin{eqnarray}
m_{N_1}U_{N_1\ell}^2 \lambda_1\quad  &\to& \quad 
m_{N_1}U_{N_1\ell}^2 \lambda_1\left( 1 + i \frac{m_{N_1} \Gamma_{12}}{\Delta m^2} \frac{U_{N_2\ell}}{U_{N_1\ell}} \frac{(\lambda_1 + \lambda_2)}{\lambda_1}\right),
\nonumber
\\
m_{N_1}U_{N_1\ell}^2 \lambda_1 \quad &\to& \quad
m_{N_2}U_{N_2\ell}^2 \lambda_2\left( 1 - i \frac{m_{N_2} \Gamma_{12}}{\Delta m^2} \frac{U_{N_1\ell}}{U_{N_2\ell}} \frac{(\lambda_1 + \lambda_2)}{\lambda_2}\right).
\end{eqnarray}
In what follows, we will disregard this correction, as it only contributes to specific numerical results, but not as a conceptual part of the analysis, and in any case it can be easily added at the end.

After simple operations, the CP asymmetry can be expressed as:
\begin{equation}
A_{CP} = \frac{2 \int d\rho_f  \  |\mathcal{M}_{N_1} \mathcal{M}_{N_2}^\ast | \sin\Delta\eta \sin\Delta\phi   }
{\int d\rho_f \ (|\mathcal{M}_{N_1}|^2+|\mathcal{M}_{N_2}|^2 + 2 |\mathcal{M}_{N_1} \mathcal{M}_{N_2}^\ast | \cos\Delta\eta \cos\Delta\phi ) } , \label{eq:ACP}
\end{equation}
where $d\rho_f$ is the final 3-body phase space, common to both the decay and its conjugate.
Clearly the asymmetry vanishes if $\Delta\phi$ is zero (modulo $\pi$), and the maximum asymmetry occurs if $\Delta\phi = \pi/2$ (or an odd factor of $\pi/2$). Notice that $\Delta\eta$, instead, is not a ``free'' parameter but it is a function of the integration variables.


In order to get a potential estimate of this asymmetry, in what follows we will assume the CP-odd phase is $\Delta \phi = \pi/2$ (maximal asymmetry). A more general case for the CP-odd phase $\Delta \phi$ can be treated in a similar way, but it will lead to longer expressions that, at this stage, will not be more illuminating.  The only caveat is that a general $\Delta\phi$ will induce smaller CP asymmetries than those inferred here. The CP asymmetry attained with $\Delta \phi = \pi/2$ is:
\begin{equation}
\hat A_{CP} = \frac{2 \int d\rho_f  \  |\mathcal{M}_{N_1} \mathcal{M}_{N_2}^\ast | \sin\Delta\eta }
{\int d\rho_f \ (|\mathcal{M}_{N_1}|^2+|\mathcal{M}_{N_2}|^2  ) } .
\label{maxACP}
\end{equation}

It is a straightforward calculation to show that the integrand in the numerator is:
\begin{eqnarray}
|\mathcal{M}_{N_1} \mathcal{M}_{N_2}^\ast | \sin\Delta\eta
&=& C_f \  |L|^2 m_K^2 \ |U_{N_1\ell} |^2  m_{N_1} \  |U_{N_2\ell} |^2 m_{N_2}
\\
&&\times \Bigg(
\frac{(p_N^2 -m_{N_1}^2)}  {(p_N^2 -m_{N_1}^2)^2 + m_{N_1}^2 \Gamma_{N_1}^2} \
\frac{  m_{N_2} \Gamma_{N_2}}{(p_N^2 -m_{N_2}^2)^2 + m_{N_2}^2 \Gamma_{N_2}^2}
\nonumber \\
&&
-\frac{   m_{N_1}\Gamma_{N_1}}{(p_N^2 -m_{N_1}^2)^2 + m_{N_1}^2 \Gamma_{N_1}^2} \
\frac{      (p_N^2 -m_{N_2}^2)}{(p_N^2 -m_{N_2}^2)^2 + m_{N_2}^2 \Gamma_{N_2}^2}\Bigg)  .
\nonumber
\end{eqnarray}

Since $\Gamma_{N_i} \ll m_{N_i}$ (they are weakly interacting) we can use the narrow-width approximation, namely
$m_N \Gamma_N/[(p_N^2 -m_{N}^2)^2 + m_{N}^2 \Gamma_{N}^2] \simeq \pi \delta(p_N^2-m_N^2)$, to reduce the previous expression to:
\begin{eqnarray}
\label{numerator1}
|\mathcal{M}_{N_1} \mathcal{M}_{N_2}^\ast | &&\sin\Delta\eta \  = \ 
C_f  \  |L|^2 m_K^2  \ |U_{N_1\ell} |^2  m_{N_1} \  \  |U_{N_2\ell} |^2  m_{N_2}
 \\
&&\times\Bigg( \frac{ \Delta m_{N}^2 }{(\Delta m_{N}^2)^2 + m_{N_1}^2 \Gamma_{N_1}^2}  \  \pi \delta (p_N^2 -m_{N_2}^2)
+
\frac{ \Delta m_{N}^2  }{(\Delta m_{N}^2 )^2 + m_{N_2}^2 \Gamma_{N_2}^2} \  \pi \delta (p_N^2 -m_{N_1}^2)   \Bigg) ,
\nonumber
\end{eqnarray}
where $\Delta m_N^2 \equiv m_{N_2}^2 - m_{N_1}^2$.  With the same narrow-width approximation, the integrand in the denominator of Eq.~(\ref{maxACP}) reduces to:
\begin{equation}
\label{denominator1}
|\mathcal{M}_{N_1}|^2+|\mathcal{M}_{N_2}|^2=
C_f \  |L|^2 m_K^2 \Bigg(
|U_{N_1 \ell}|^4
\frac{ m_{N_1}}{\Gamma_{N_1}} \ \pi\delta (p_N^2-m_{N_1}^2)
+
|U_{N_2 \ell}|^4  \frac{m_{N_2}}{\Gamma_{N_2}}\  \pi \delta (p_N^2-m_{N_2}^2) \Bigg) .
\end{equation}

Now, for the 3-body phase space integration $\int d\rho_f$ of this expression, we should notice that the only factors that depend on the integration variables are the delta function
and the lepton trace $|L|^2$, so we only need to solve the dimensionless integral:
\begin{eqnarray}
{\cal I}(m_N)\equiv \int d\rho_f \ \pi\delta (p_N^2-m_{N}^2) \ |L|^2   .
\label{Theintegral}
\end{eqnarray}
The explicit expression for ${\cal I}(m_N)$ is given in Appendix A, and is a smooth function of $m_N$.

Moreover, for  almost degenerate neutrinos ($m_{N_1}\simeq m_{N_2}$) the integrals ${\cal I}(m_{N_1})$ and
${\cal I}(m_{N_2})$  are almost the same, so they cancel in the ratio and the expression for $\hat A_{CP}$ greatly simplifies:
\begin{equation}
\hat A_{CP} \simeq
\frac{
2  \ |U_{N_1\ell} |^2  m_{N_1} \  \  |U_{N_2\ell} |^2  m_{N_2}
 }
{
\Bigg(
|U_{N_1 \ell}|^4
{ m_{N_1}}/{\Gamma_{N_1}}
 +
|U_{N_2 \ell}|^4  m_{N_2}/\Gamma_{N_2}
 \Bigg)
}\Bigg(  \frac{ \Delta m_{N}^2 }{(\Delta m_{N}^2)^2 + m_{N_1}^2 \Gamma_{N_1}^2}
 +
\frac{ \Delta m_{N}^2  }{(\Delta m_{N}^2 )^2 + m_{N_2}^2 \Gamma_{N_2}^2}
\Bigg) .
\end{equation}


From this expression, it is evident that the asymmetry is suppressed if the sterile neutrinos
are exactly degenerate ($\Delta m_N^2 \to 0$), but it is also suppressed as the mass splitting gets too large, namely as $\Delta m_N^2 \gg m_N \Gamma_N$.

\begin{figure}[htb]
\begin{center}
\includegraphics[scale=1]{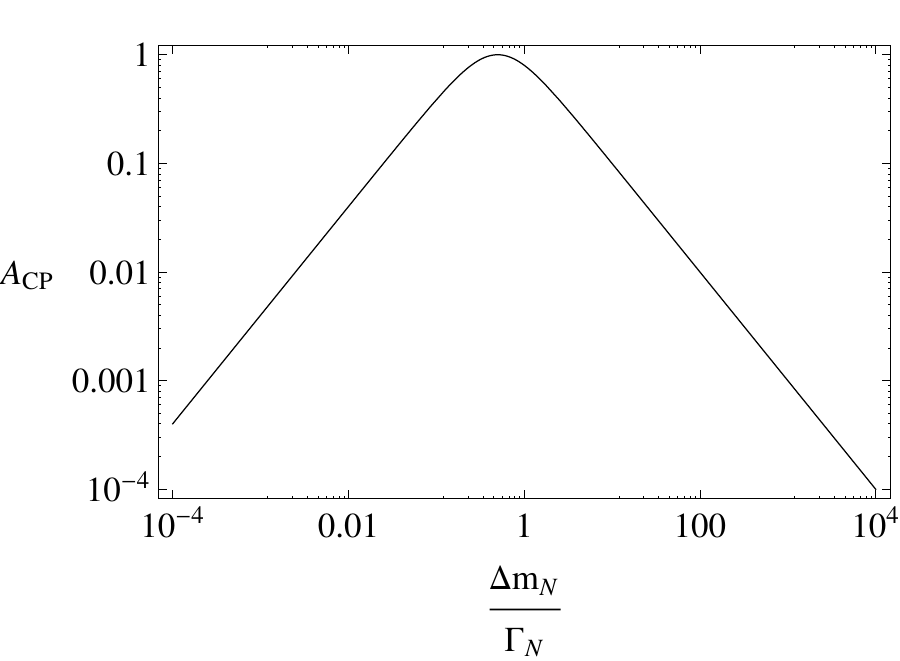}
\end{center}
\caption{The CP asymmetry of Eq.~(\ref{eq:ACPsimp}), for values of the heavy neutrino mass splitting $\Delta m_N$ near their decay width $\Gamma_N$. }
\label{ACPfig}
\end{figure}

As a crude estimate, we can assume that the lepton mixing elements are comparable (albeit small), $|U_{N_1 \ell}| \sim |U_{N_2 \ell}|$, and also the decay rates, $\Gamma_{N_1} \sim \Gamma_{N_2}$. Then the asymmetry simplifies to:
\begin{equation}
\hat A_{CP} \sim
2   m_{N} {\Gamma_{N}}
\frac{ \Delta m_{N}^2 }{(\Delta m_{N}^2)^2 + m_{N}^2 \Gamma_{N}^2} .
 \label{eq:ACPsimp}
\end{equation}
Under these assumptions the mixing elements $U_{N\ell}|^2$ cause the suppression of the branching ratios but not of the CP asymmetry. The latter is maximal ($\sim 1$) when
\begin{equation}
\Delta m_N^2 \sim m_N \Gamma_N,\quad i.e.\ \ \Delta m_N \sim \Gamma_N/2 \label{eq:Dm}
\end{equation}
and becomes smaller when $\Delta m_N$  and $\Gamma_N$ are not of comparable size (either as $\Delta m_N \ll \Gamma_N$ or $\Delta m_N \gg \Gamma_N$). Fig.~\ref{ACPfig} shows this behavior for values of $\Delta m_N$ near $\Gamma_N$. Here we have defined \hbox{$\Delta m_N \equiv m_{N_2} - m_{N_1}$}, not to be confused with $\Delta m_N^2 \equiv m_{N_2}^2 - m_{N_1}^2$.

\begin{figure}[b]
\begin{center}
\includegraphics[scale=1]{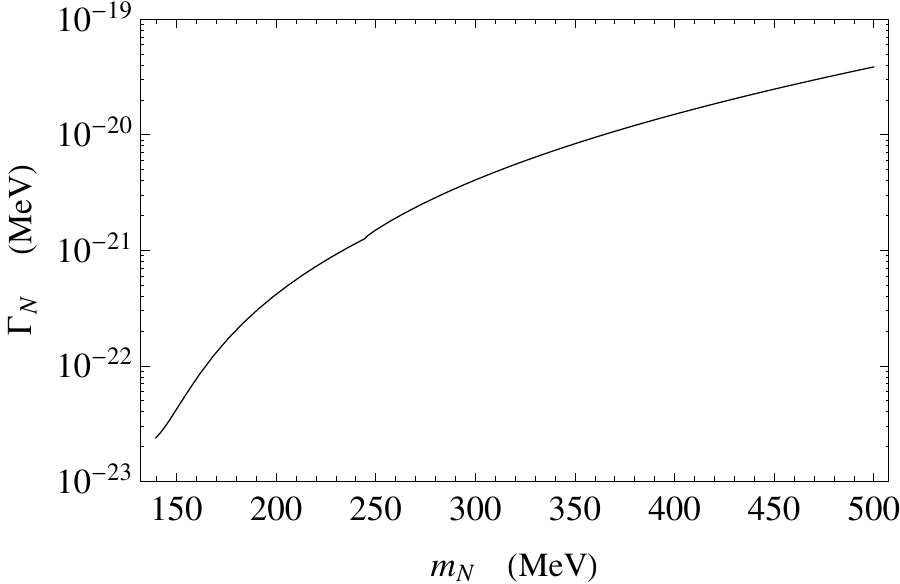}
\end{center}
\caption{The total decay width of a sterile neutrino $N$ as function of its mass, in the range between $m_\pi$ and $M_K$, considering the lepton mixing elements of the order $|U_{N\ell}|^2 \sim 10^{-9}$, a value that is representative of current bounds.}
\label{fig:GammaN}
\end{figure}

To complete the estimate of the CP asymmetry we need to calculate the neutrino decay rates
$\Gamma_{N_1}$ and $\Gamma_{N_2}$ within our framework. For the decay of a Majorana neutrino $N$ with a mass between $m_\pi+m_\ell$ and $m_K-m_\ell$, the relevant channels are $\Gamma_N\to \ell^\pm\pi^\mp$, $\nu_\alpha\pi^0$,
$\ell^\pm \ell^{\prime\mp}\nu_{\ell^\prime}$, $\nu_\alpha\ell^+\ell^-$, $\nu_\alpha \nu_{\alpha^\prime}\bar\nu_{\alpha^\prime}$, where $\ell, \ell^\prime = e, \mu$, and $\alpha, \alpha^\prime = e,\mu,\tau$.
The expressions for these rates were given in Ref.~\cite{Atre:2009rg}.
In order to get numerical values for the width, we need estimates for the lepton mixing elements. As a crude estimate, we use here the conservative values $|U_{N\ell}|^2\approx 10^{-9}$ ($\ell=e,\mu, \tau$), according to current bounds \cite{Atre:2009rg,Ruchayskiy:2011aa}.
The result for the decay width
$\Gamma_N$ as a function of $m_N$  is shown in Fig.~\ref{fig:GammaN}.
As can be seen in the figure, the predicted decay width of the sterile neutrino is a growing function of its mass. This is so because each partial width is itself a growing function of the mass, either as $m_N^3$ (two-body decays) or as $m_N^5$ (three-body decays), and in addition more decaying channels are opening as $m_N$ increases.
The resulting widths $\Gamma_N$ vary from  about  $2\times 10^{-23}$ MeV (for $m_N \sim 140$ MeV) to about $  4 \times 10^{-20}$ MeV (for $m_N \sim 500$ MeV), corresponding to lifetimes between $10^1$ sec and $10^{-2}$ sec, respectively.
These lifetimes are very long, so that in a general case the two vertices where the charged leptons are produced will be far apart, more so if the decaying kaons have large velocities, so that most of the secondary leptons will be produced outside the detector.
This feature will not affect the CP asymmetry but it could greatly reduce the number of events that can actually be detected \cite{Dib:2014iga}.

\section{Decay rates and Vertex displacements.}
\label{sec:Results}

The observability of the CP asymmetry depends not only on the size of the asymmetry itself but also on the size of the decay rates. Let us then estimate the branching ratios of these rare modes $K^\pm \to \pi^\mp \ell^\pm\ell^\pm$, considering current experimental limits.
The expression for the branching ratios in the scenario $\Delta\phi= \pi /2$ is:
\begin{equation}
Br (K^\pm\to \pi^\mp \ell^\pm\ell^\pm )=
\frac{1}{2 m_K  \Gamma_K}    \int d\rho_f \left(   |\mathcal{M}_{N_1}|^2  +  |\mathcal{M}_{N_2}|^2  \pm 2  |\mathcal{M}_{N_1} \mathcal{M}_{N_2}| \sin\Delta\eta  \right)  .
\end{equation}

The third term in this expression is the interference between the two intermediate neutrinos and is the cause of the CP asymmetry. It  is at most equal to the previous two terms in the particular case of Eq.~(\ref{eq:Dm}), or much less if the neutrinos are far from degenerate. To have a general estimate of the branching ratios we will consider the \emph{average} of the mutually conjugate modes:
\begin{equation}
\overline{Br}(K\to\pi\ell\ell) \equiv \frac{1}{2} \left\{Br(K^+\to\pi^-\ell^+\ell^+) +  Br(K^-\to\pi^+\ell^-\ell^-)\right\},
\end{equation}
which depends on $ |\mathcal{M}_{N_1}|^2  +  |\mathcal{M}_{N_2}|^2 $ only, not on the interference.  An additional issue we should clarify concerns the
crossed diagrams in the case of identical charged leptons in the final state. Since the intermediate neutrinos cannot go on mass shell at the same kinematical point, the crossed diagrams do not interfere. As a result, to consider the crossed diagrams and the symmetry factor $1/2$
is equivalent to considering no crossed diagrams at all. Moreover, for almost degenerate neutrinos, which is our case of interest, it is enough to consider
$ |\mathcal{M}_{N_1}|^2  \simeq  |\mathcal{M}_{N_2}|^2 $. With these considerations, using Eqs.~(\ref{denominator1}) and (\ref{Theintegral}), the average branching ratio is:
\begin{equation}
\overline{Br} (K \to \pi  \ell \ell  ) =
G_F^4  f_\pi ^2 f_K ^2 m_K^4 \   |V_{ud}|^2 |V _{us}|^2
|U_{N \ell}|^4  \frac{m_K}{ \Gamma_K}
\frac{ m_{N}}{\Gamma_{N}} {\cal I}(m_{N})  .
\label{averageBr}
\end{equation}

\begin{figure}[t]
\begin{center}
\includegraphics[scale=0.75]{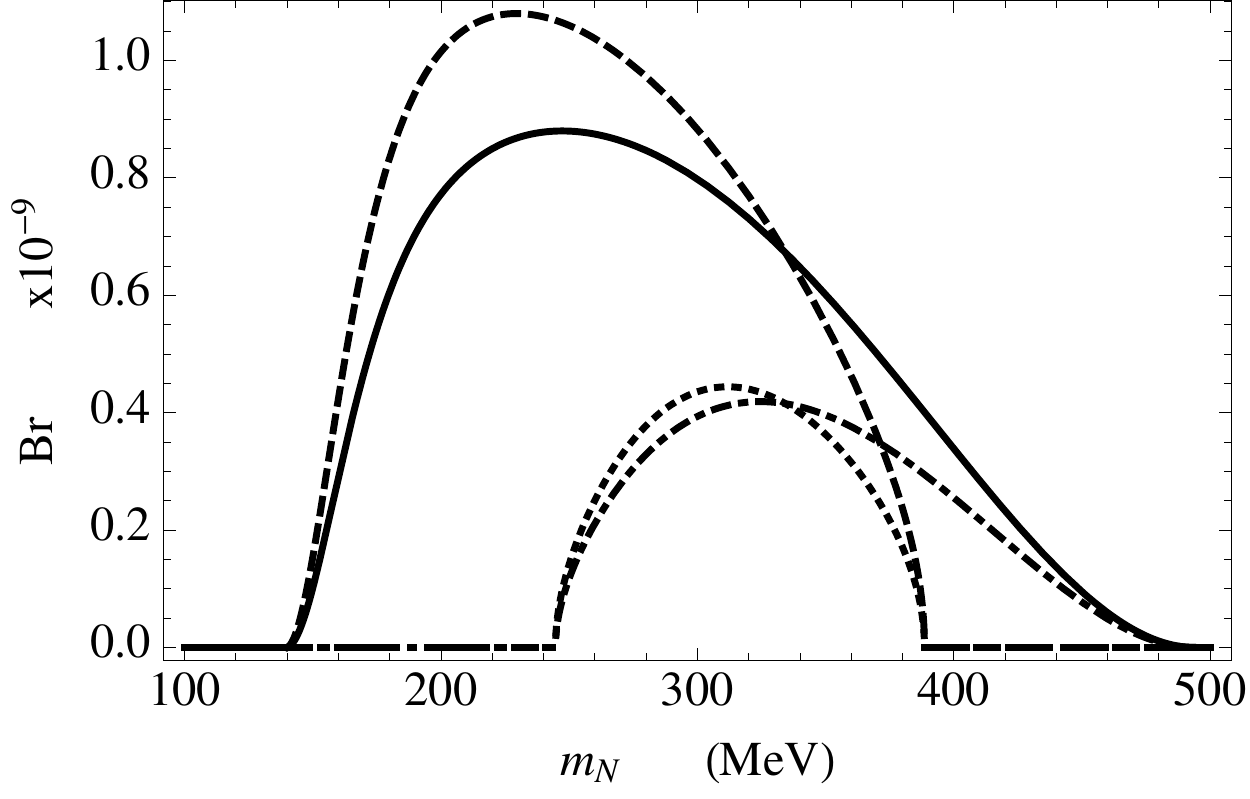}
\end{center}
	\caption{Average branching ratios (\textit{c.f.}  Eq.\ \ref{averageBr}) for  $K\to \pi e e$ (solid), $K \to \pi \mu  e $ (dashed), $K \to \pi  e \mu$ (dot-dashed),
	$K \to \pi \mu  \mu $ (dotted)   as a function of the intermediate neutrino mass $m_N$, assuming lepton mixing $|U_{N \ell }|^2 = 10^{-9}$ for all three lepton flavors.
}
	\label{fig:BrK}
\end{figure}

This expression depends on the charged lepton masses through the integral ${\cal I}(m_N)$ (see Appendix A).
For different final leptons, the allowed mass range of $m_N$ is also different. We can distinguish four cases for
$\ell_1^\pm \ell_2^\pm$, namely   $e^\pm e^\pm$, $e^\pm\mu^\pm$, $\mu^\pm e^\pm$ and $\mu^\pm \mu^\pm$.
The branching ratios for each of these cases, with the assumption that all the lepton mixing elements with the sterile neutrinos $N_1$ and $N_2$ are $|U_{N\ell}|^2 \sim 10^{-9}$, are shown in Fig. \ref{fig:BrK}.
%
%
%
Given that the current experimental upper bounds on these branching ratios are near
$10^{-9} - 10^{-10}$ \cite{Beringer:1900zz},
the graphs show that these processes would be within the reach of current experiments provided the mixings are near the current bounds and, most of all, that the whole final state is observed. The latter provision, however, is hard to fulfil due to the rather long lifetime of sterile neutrinos with mass in the range of interest. Unless one has an extremely large detector, most of the second vertices (i.e. the decay $N\to \pi \ell$) will occur outside the detector, thus losing  the information on the Majorana character of the neutrino and on the CP asymmetry.
Because of this phenomenon of \emph{vertex displacement}, the observability of these processes is highly reduced \cite{Dib:2014iga}.

For neutrinos produced in average with velocity $\beta_N$ (using $c=1$) and Lorentz factor $\gamma_N$, the characteristic distance they travel before decaying is $L_N = \gamma_N \beta_N \tau_N$, where $\tau_N = 1/\Gamma_N$ is the sterile neutrino lifetime, which depends on the neutrino mass, $m_N$.  For $N$ produced in kaon decays, \emph{i.e.} $m_N < m_K-m_\ell$, $L_N$ is a very long distance. In general, the fraction of the $N$'s that decay inside a detector of length
$L_D$ (usually $L_D\ll L_N$) is:
\begin{equation}
P_N  \equiv 1 - \exp(L_D/L_N) \quad \approx \frac{L_D}{L_N}.
\end{equation}
$P_N$ is then an \emph{acceptance factor} for the $\Delta L=2$ decays $K\to \pi\ell\ell$.
Let us make some estimates. Assuming all lepton mixings to be $|U_{N\ell}|^2\sim 10^{-9}$ as  in Fig. \ref{fig:GammaN}, the lifetime $\tau_N$ varies between 1 [s] (for $m_N\sim 150$ MeV)  and 10 [ms] (for $m_N\sim 500$ MeV). The decay length $L_N$ then ranges from $\gamma_N \beta_N \times 10^8$ [m] to $\gamma_N \beta_N \times 10^6$ [m], respectively.


If we consider the kaons decaying at rest (or nearly at rest), the produced neutrinos $N$ will have $\gamma_N \beta_N \sim {\cal O}(1)$ (disregarding   the threshold case $m_N \to m_K-m_\ell$, where $\beta_N\to 0$).
Consequently, for a detector 10 [m] long, $P_N \approx L_D/L_N \sim 10^{-7}$ for $m_N\sim 150$ MeV, or  $P_N\sim 10^{-5}$ for $m_N\sim 500$ MeV.

If instead we consider kaons decaying in flight with energies near $ 5$ GeV, then $\gamma_N\sim 10$, and detectors 10 times longer are required in order to have the same acceptances previously estimated.

Let us now focus on the CP asymmetry, in the scenario that lead to Eq. (\ref{eq:ACPsimp}). As shown in this expression, the asymmetry could be close to unity in the restricted range of mass splittings $\Delta m_N \sim \Gamma_N/2$. Indeed, as the decay widths $\Gamma_N$ are around $10^{-22}$ to $10^{-19}$ MeV for absolute masses $m_N$ in the range from 150 to 500 MeV, respectively (see Fig.~\ref{fig:GammaN}), the asymmetry will be sizeable only if the mass splitting is near:
\begin{equation}
\frac{\Delta m_N}{m_N} \sim 10^{-24} - 10^{-22},
\end{equation}
for $m_N \sim 150 - 500$ MeV, respectively. Such tiny splittings could only be detected through interference measurements  such as this asymmetry. If the actual value of $\Delta m_N$, turns out to be larger than $\Gamma_N/2$, the asymmetry decreases approximately as $A_{CP} \sim \Gamma_N / \Delta m_N $, as shown in Fig.~\ref{ACPfig}.

We must also point out that the expression for the asymmetry shown in Eq.~(\ref{eq:ACPsimp}) does not depend on the lepton flavours in the final state ($ee$, $\mu e$ or $\mu \mu$), but this is only due to our assumption that the CP-odd phase
is at its optimal value for the asymmetry, namely $\Delta \phi = \pi/2$. In nature, this condition may approximately apply to one specific channel and not another, an issue that can be resolved with further knowledge of the CP phases in the lepton sector.


\section{Summary}

We have studied the possibility to observe the $\Delta L=2$ decays $K^\pm \to \pi^\mp \ell^\pm \ell^{\prime\pm}$
mediated by sterile Majorana neutrinos with masses in the range that allow them to be on mass shell in the s-channel amplitude. In particular, we look for the conditions that allow a sizable CP violation in the conjugate rates, and found that this is the case if there are two or more neutrinos almost degenerate in mass, with mass splitting comparable to their decay rate, $\Delta m_N \sim \Gamma_N/2$. However, for neutrinos with masses below the kaon mass, as required, current estimates of the decay rates indicate very long lifetimes, and consequently in an experimental search one should consider that only a very small fraction of the decays may occur fully inside the detector. This particular suppression is less severe in decays of mesons heavier than kaons, as they require intermediate neutrinos with larger masses and consequently shorter lifetimes.







\acknowledgments
This work was supported in part by Conicyt (Chile) grant \textit{Institute for Advanced Studies in Science and Technology} ACT-119, by Fondecyt (Chile) grant 1130617, by the National Research Foundation of Korea (NRF)
grant funded by Korea government of the Ministry of Education, Science and
Technology (MEST) (No. 2011-0017430) and (No. 2011-0020333).
\\

\appendix

\section{Calculation of the decay rate integrals.} \label{App:DC}

The decay rate $K^+ \to \ell^+ \ell^+ \pi^-$ mediated by a sterile neutrino $N$ in the intermediate state according to Eq.\ (\ref{amplitude1}), requires an integral over the 3-body phase space of the form
[see Eqs.\ (\ref{numerator1}) and  (\ref{denominator1})]:
\begin{eqnarray}
{\cal I}(m_N) \equiv
\int d\rho_f^{(3)} \pi\delta (p_N^2-m_{N}^2)\   |L|^2 .
\label{theintegral}
\end{eqnarray}
All other factors that appear in the transition probability are constant with respect to the integration, so we did not include them in the definition of ${\cal I}(m_N)$. Here $|L|^2$ is the lepton trace given in Eq.\ (\ref{trace}), and the 3-body phase space in our case is
\[
\int d\rho_f^{(3)} = \int \frac{d^3\ell_1}{(2\pi)^3 2E_1}
 \frac{d^3 \ell_2}{(2\pi)^3 2E_2} \frac{d^3 p}{(2\pi)^3 2E_p} (2\pi)^4 \delta^4 (k-\ell_1 -\ell_2 -p) ,
\]
where we denote $k$ the momentum of the decaying kaon, while $\ell_1$, $\ell_2$ and $p$ are the momenta of the final particles (two leptons and a pion, respectively). The 4-momentum $p_N$ that
appears in the delta function is by definition $p_N = \ell_2 + p$, and corresponds to the momentum of the sterile neutrino $N$ in the intermediate state.

As it is well known, the 3-body phase space, i.e. the Dalitz plot, reduces to the integration over two energies. Here the delta function in the integrand further reduces the integral over a single energy, which we choose it to be the pion energy in the kaon rest frame (normalised to the kaon mass), $\varepsilon_p = E_p/m_K$:
\begin{eqnarray}
\int d\rho_f^{(3)} \pi\delta (p_N^2-m_{N_i}^2) =
\frac{1}{ 64 \pi^2  }
 \int
 \ d\varepsilon_p ,
\end{eqnarray}

Now, concerning the integrand of Eq.\ (\ref{theintegral}), it is straightforward to show that  $|L|^2$ can be expressed in terms of
$\varepsilon_p$ as:
\begin{equation}
|L|^2 = 4 A - 4 B\  \varepsilon_p ,
\end{equation}
where $A$ and $B$ are the following constants:
\begin{eqnarray}
A
&=&    (x_N^2 - x_2^2)^2 +   x_\pi^2 (x_N^2 - x_1^2)^2 -  x_\pi^2 (x_1^2 + x_2^2)  ,
\nonumber \\
B
&=&  2\,  (x_N^2 - x_1^2) (x_N^2 - x_2^2)  ,\label{eq:AB}
\end{eqnarray}
and $x_i \equiv m_i/m_K$ are the particle masses in units of $m_K$.
Now, concerning the integration limits of the pion energy $\varepsilon_p$, they are determined by energy-momentum conservation and  by the condition that the intermediate neutrino $N$ must be on mass shell. These limits can be expressed in a simple way as follows:
\begin{equation}
\varepsilon_p^{\ max/ min} = \frac{\varepsilon_N \ \varepsilon_p^{(0)}}{x_N} \pm \frac{\textrm{p}_N\  \textrm{p}^{(0)}} {x_N} ,
\end{equation}
where $\varepsilon_N$ and $\textrm{p}_N$ are the energy and momentum of the intermediate neutrino $N$ in the
kaon rest frame (in this frame, $N$ is mono energetic):
\begin{equation}
\varepsilon_N = \frac{1 + x_N^2 -x_1^2}{2},
\quad
\textrm{p}_N = \sqrt{\varepsilon_N^2 - x_N^2} ,
\label{eq:EpN}
\end{equation}
while $\varepsilon_p^{(0)}$ and  $\textrm{p}^{(0)}$ are the pion energy and momentum in the $N$ rest frame (likewise, in this frame the pion is mono energetic):
\begin{equation}
\varepsilon_p^{(0)} = \frac{x_N^2 + x_\pi^2 -x_2^2}{2 x_N},
\quad
\textrm{p}^{(0)} = \sqrt{ \varepsilon_p^{(0)2} - x_\pi^2} .
\label{eq:Ep0}
\end{equation}

Bringing all of the above together, the integral in Eq. (\ref{theintegral}) results in a rather compact expression:
\begin{eqnarray}
{\cal I}(m_N) &\equiv & \int d\rho_f^{(3)} \pi\delta (p_N^2-m_{N_i}^2) |L|^2
\nonumber \\
&=&
 \frac{1}{8\pi^2 } \frac{\textrm{p}_N\ \textrm{p}^{(0)} }{ x_N}  \left(  A  -    B \ \frac{  \varepsilon_N\ \varepsilon_p^{(0)} }{x_N} \right) ,
\end{eqnarray}
with $\textrm{p}_N$, $\textrm{p}^{(0)}$, $\varepsilon_N$, $\varepsilon_p^{(0)}$, $A$ and $B$ as given above. The integral
${\cal I}(m_N)$ is a smooth function of the intermediate neutrino mass $m_N$.
\\


\end{document}